\documentclass[twocolumn,showpacs,preprintnumbers,amsmath,amssymb]{revtex4}

\usepackage{graphicx}
\usepackage{dcolumn}
\usepackage{bm}

\begin{document}

\preprint{APS/123-QED}

\title{Equality of Bond Percolation Critical Exponents for Pairs of Dual Lattices}

\author{Matthew R. A. Sedlock}
 \altaffiliation{Research supported under the Acheson J. Duncan Fund
 in the Applied Mathematics and Statistics Department of The Johns Hopkins University.}
 \affiliation{Applied Mathematics and Statistics Department,
 The Johns Hopkins University, Baltimore, Maryland 21218}

\author{John C. Wierman}
 \altaffiliation{Research supported under the Acheson J. Duncan Fund
 in the Applied Mathematics and Statistics Department of The Johns Hopkins University.}
 \affiliation{Applied Mathematics and Statistics Department,
 The Johns Hopkins University, Baltimore, Maryland 21218}


\begin{abstract}
For a certain class of two-dimensional lattices, lattice-dual pairs
are shown to have the same bond percolation critical exponents.  A
computational proof is given for the martini lattice and its dual to
illustrate the method. The result is generalized to a class of
lattices that allows the equality of bond percolation critical
exponents for lattice-dual pairs to be concluded without performing
the computations.  The proof uses the substitution method, which
involves stochastic ordering of probability measures on partially
ordered sets. As a consequence, there is an infinite collection of
infinite sets of two-dimensional lattices, such that all lattices in
a set have the same critical exponents.
\end{abstract}


\maketitle

\section{\label{sec:level1}Introduction}

The behavior of a percolation system near the critical probability
is often expressed in terms of critical exponents. Their values are
believed to depend only on the dimension of the lattice, rather than
the structure of the lattice itself.  This is known as the
universality hypothesis.  Using scaling and hyperscaling relations,
values for the critical exponents in two dimensions have been
proposed and supported by substantial simulation evidence
\cite{Hug96}.

The works of Kesten \cite{Kes87} and Wierman \cite{Wie92} give
mathematical progress toward showing the universality hypothesis in
two dimensions; Kesten was able to show relationships among critical
exponents assuming that the limits defining them exist, while
Wierman was able to establish the equality of certain bond
percolation critical exponents for two pairs of dual lattices: the
triangular-hexagonal pair and the bowtie-dual pair.

Very little has been proved about critical exponents in two
dimensions.  Kesten \cite{Kes82} shows that specific functions can
be bounded above and below by powers of $|p-p_c|$ for lattices in
which the ``horizontal and vertical direction play symmetric
roles.''  In particular, these results do not rely on the
calculation of the percolation threshold, $p_c$, of a given lattice,
but rely instead on other properties of the lattice.  Wierman's
result for the triangular-hexagonal pair and the bowtie-dual pair
rely heavily on the ability to calculate the percolation threshold
for these lattices using the star-triangle transformation.  As a
result, little use has been made of this method since the exact
percolation threshold is known for relatively few lattices.  A
notable advance by Smirnov and Werner \cite{SmWe} is the
determination of the values of the critical exponents of the site
percolation model on the triangular lattice.

The recent work of Ziff and Scullard \cite{Zif06,Zif062,Scu06}
generalizes the star-triangle transformation to find the exact
percolation thresholds for the martini lattice and its variants, the
A lattice and the B lattice.  Using this idea, Wierman and Ziff
\cite{Wie08} have since identified an infinite class of
lattices--those constructed from an infinite connected planar
periodic 3-uniform hypergraph with one axis of symmetry, using a
generator which is a finite connected planar graph with three
boundary vertices--for which the exact bond percolation threshold
can be determined.  By applying Wierman's method to the martini
lattice and its dual, this paper gives a third lattice-dual pair
with the same bond percolation critical exponents. More importantly,
the method can be extended to the class identified by Wierman and
Ziff, so that infinitely many lattice-dual pairs are shown to have
the same values for their critical exponents. Although the ability
to calculate the percolation threshold is essential for proving the
result for lattices in the class mentioned, the result does not
depend on the percolation threshold of a lattice.

Section II provides some background information, describes the class
of lattices, and explains the method.  Section III uses the method
of Wierman \cite{Wie92} to show equality of the critical exponents
for the martini lattice and its dual.  Section IV shows that the
method of Wierman can be extended to lattices constructed from an
infinite connected planar periodic 3-uniform hypergraph with one
axis of symmetry, using a generator which is a finite connected
planar graph with three boundary vertices. Thus, without actually
going through the calculations, we can conclude that a lattice
belonging to this class and its dual have the same bond percolation
critical exponents. Section V shows that the result includes other
lattices that can be transformed into lattices in the class and
discusses the implications for the equivalent site problems.

\section{\label{sec:level1}Preliminaries}

\subsection{\label{sec:level2}Percolation Functions and Critical Exponents}

In the bond percolation model on an underlying graph $G$, bonds
between vertices are labeled open with probability $p$, $0\leq p
\leq 1$, independent of all other bonds.  A bond that is not open is
said to be closed.  The probability measure and expectation operator
corresponding to a given $p$ are $P_p$ and $E_p$ respectively.
Several interesting quantities arise in this model.  For any given
vertex $v$ in $G$, let $C_v$ be the set of bonds in the connected
component containing $v$.  Then, $|C_v|$ gives the number of bonds
in the connected component containing $v$.  The main quantities
considered in this paper are the percolation probability functions,
$\theta_v(p) = P_p[|C_v|=+\infty]$, and the mean finite cluster
size, $\chi _v(p) = E_p[|C_v|;|C_v|<\infty]$.  The two-point
connectivity function and the correlation length are also of
interest and are defined in section V.

It is believed, though not proved, that most quantities of interest
in percolation theory behave as powers of $|p-p_c|$ as $p$
approaches $p_c$. These powers are called {\it critical exponents}.
The notation
$$A(p) \approx |p-p_c|^\zeta$$
 is used to mean that
$$\mathop {\lim }\limits_{p \to p_c} \frac{\log A(p)}{\log |p-p_c|} = \zeta.$$
It is not known that these limits exist, so we use the superscripts
$+$ and $-$ on the exponent to denote limsup and liminf
respectively. The power laws we consider are
$$\theta(p) \approx (p-p_c)^\beta, \mbox{ for }p>p_c,$$
for some $0<\beta<1$, and
$$\chi_f(p) \approx |p-p_c|^{-\gamma}$$
for some $\gamma>0.$ The power laws for the two-point connectivity
function, the correlation length, and other quantities are given in
section V.

\subsection{The Class of Lattices}

We now describe the class of lattices, introduced by Wierman and
Ziff \cite{Wie08}, for which our results are valid.  We will refer
to it as the ``martini class" of lattices. The lattices are
constructed by placing copies of a planar graph called a generator
into a self-dual 3-uniform hypergraph arrangement. The concepts of
{\it generator}, {\it 3-uniform hypergraph}, and {\it self-duality}
in this context are explained in the following subsections.

\subsubsection{Self-dual Hypergraph Arrangements}

The concept of hypergraph is a generalization of the concept of
graph, in which hyperedges connect sets of vertices rather than
edges connecting pairs of vertices.  Given a set $V$ of vertices, a
hyperedge $H$ is a subset of $V$. A hyperedge $H$ is said to be
incident to each of its vertices.  A hyperedge containing exactly
$k$ vertices is called a $k$-hyperedge.

 A hypergraph is a vertex set $V$ together with a set of hyperdeges
 of vertices in $V$.  A hypergraph containing only $k$-hyperedges is a $k$-uniform
 hypergraph. A hypergraph is planar if it
 can be embedded in the plane with each hyperedge represented by a
 bounded region enclosed by a simple closed curve with its vertices
 on the boundary, such that the intersection of two hyperedges is a
 set of vertices.

In this article, we will consider only $3$-uniform hypergraphs. For
convenient visualization, we will represent each 3-hyperedge in the
plane as a shaded triangular region bounded by a slightly concave
triangular boundary. This allows us to neglect the detailed
structure of our generators when we consider arranging them in the
plane to form a connected structure.

In order to construct exactly-solvable lattice graphs for bond
percolation models, we will consider infinite connected planar
periodic 3-uniform hypergraphs. A planar hypergraph $H$ is periodic
if there exists an embedding with a pair of basis vectors {\bf u}
and {\bf v} such that $H$ is invariant under translation by any
integer linear combination of {\bf u} and {\bf v}, and such that
every compact set of the plane is intersected by only finitely many
hyperedges.

If a hypergraph $H$ is planar, we may construct its dual hypergraph
$H^*$ as follows. Place a vertex of $H^*$ in each face of $H$.  For
each hyperedge $e$ of $H$, construct a hyperedge $e^*$ of $H^*$
consisting of the vertices in the faces surrounding $e$. Note that
if the hyperedge of $H$ is a 3-hyperedge represented by a triangular
region, and each of its boundary vertices is in at least two
hyperedges, then the dual hyperedge is a 3-hyperedge also,
represented by a ``reversed triangle."

Two hypergraphs are isomorphic if there is a one-to-one
correspondence between their vertex sets which preserves all
hyperedges.  A hypergraph is self-dual if it is isomorphic to its
dual.  If, in addition, the hypergraph is 3-uniform, this
corresponds to the term triangle-dual used by Ziff and Scullard.  To
illustrate, Figure 1 provides two examples of infinite connected
planar periodic self-dual 3-uniform hypergraphs mentioned by Ziff
and Scullard.

\begin{figure}
\includegraphics[scale=.40]{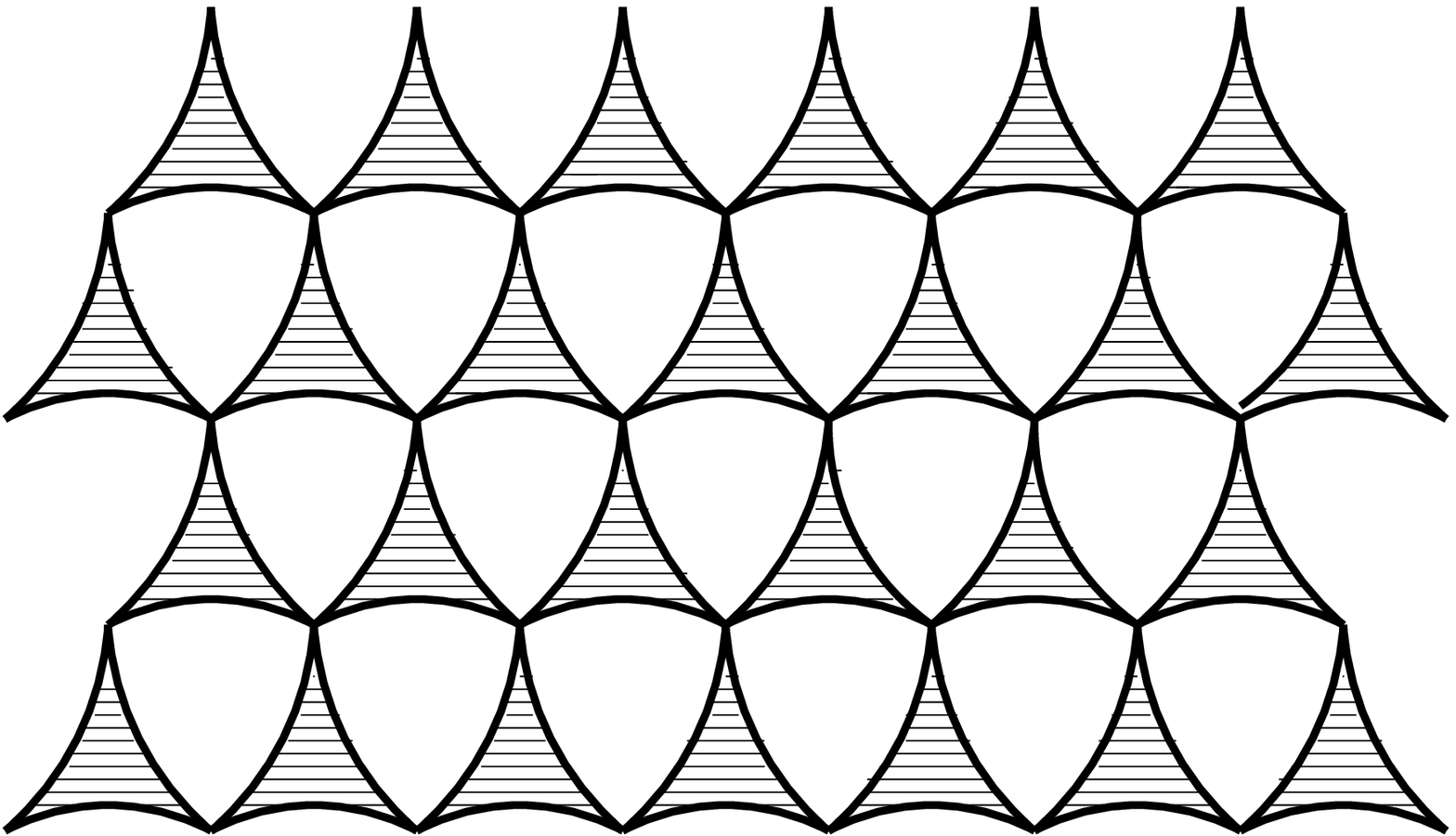}

\includegraphics[scale=.40]{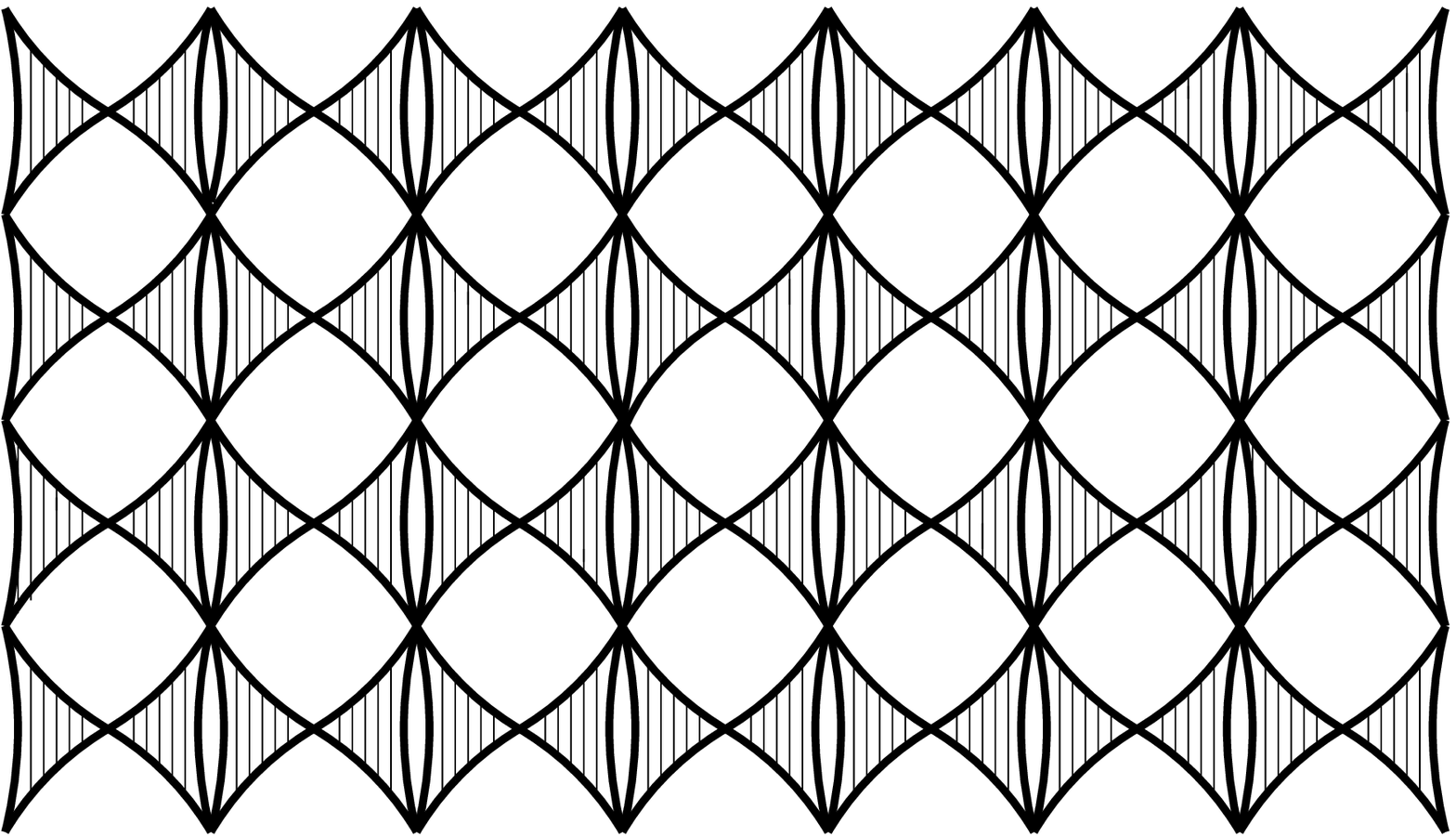}
\caption{Self-dual hypergraph arrangements illustrated in Ziff and
Scullard. We refer to the upper drawing as the triangular
arrangement and refer to the lower drawing as the bow-tie
arrangement. }
\end{figure} \label{self-dual-arrangements}

Wierman and Ziff \cite{Wie08} note that the reversed triangles must
be connected in a specific manner to create a self-dual arrangement,
and provide an example to illustrate that simply reversing each
triangle is not sufficient.

\subsubsection{Generators and Duality}

A generator is a finite connected planar graph embedded in the plane
so that three vertices on the infinite face are designated as
boundary vertices A, B, and C. (For future research, the term
generator may be defined more generally. However, in this paper, we
restrict attention to generators with three boundary vertices.)

Given a generator $G$, we construct its dual generator $G^*$ by
placing a vertex in each bounded face of $G$, and three vertices
$A^*$, $B^*$, and $C^*$ of $G^*$ in the infinite face of $G$, as
follows: The boundary of the infinite face can be decomposed into
three (possibly intersecting) paths, from $A$ to $B$, $B$ to $C$,
and $C$ to $A$. The infinite face may be partitioned into three
infinite regions by three non-intersecting polygonal lines starting
from $A$, $B$, and $C$.  Place $A^*$ in the region containing the
boundary path connecting $B$ and $C$, $B^*$ in the region containing
the boundary path connecting $A$ and $C$, and $C^*$ in the region
containing the boundary path connecting $A$ and $B$. $A^*$, $B^*$,
and $C^*$ are the boundary vertices of $G^*$.

For each edge $e$ of $G$, construct an edge $e^*$ of $G^*$ which
crosses $e$ and connects the vertices in the faces on opposite sides
of $e$. If $e$ is on the boundary of the infinite face, connect it
to $A^*$ if $e$ is on the boundary path between $B$ and $C$, to
$B^*$ if $e$ is between $A$ and $C$, and connect it to $C^*$ if $e$
is between $A$ and $B$. (Note that it is possible for $e^*$ to
connect more than one of $A^*$, $B^*$, and $C^*$, for example, if
there is a single edge incident to $A$ in $G$, its dual edge
connects $B^*$ and $C^*$.)

\begin{figure}
\includegraphics[scale=.5]{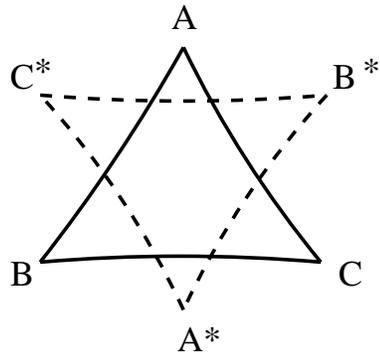} \caption{Solid
lines represent a 3-hyperedge with boundary vertices $A$, $B$, and
$C$. Dashed lines represent the ``reversed" or dual hyperedge, with
its boundary vertices $A^*$, $B^*$, and $C^*$ labeled in the proper
positions.  We say that $A^*$ is opposite $A$, $B^*$ is opposite
$B$, and $C^*$ is opposite $C$.}
\end{figure} \label{reversed-triangle}

Note that $G^*$ is not the dual graph of $G$, which would have only
one vertex in the infinite face.  The three vertices $A^*$, $B^*$,
and $C^*$ will correspond to separate faces of the lattice generated
from $G$.

\subsubsection{Constructing a Dual Pair of Lattices}

Given a planar generator $G$ and a connected periodic self-dual
3-uniform hypergraph ${\cal H}$, a dual pair of periodic lattices
may be constructed as follows:  Construct a lattice graph $L_{G,
{\cal H}}$ by replacing each hyperedge of {\cal H} by a copy of the
generator $G$, with the boundary vertices of the generator
corresponding to the vertices of the hyperedge, in such a manner
that the resulting lattice is periodic.  This is always possible, by
choosing the orientations of the generator in one period of the
hypergraph, and extending the choice periodically.  (However, for a
generator without sufficient symmetry, it may be possible to replace
hyperedges in a way that produces a non-periodic lattice, so some
care is needed.)

We now construct a lattice $L_{G^*, {\cal H}^*}$ as follows:
Construct the embedding of the dual hypergraph ${\cal H}^*$ in the
plane, in which every hyperedge of ${\cal H}$ is reversed.  Replace
each hyperedge of ${\cal H}^*$ by a copy of the dual generator
$G^*$, embedded so that it is consistent with the embedding of $G$,
that is, in all hyperedges boundary vertex $A^*$ in $G^*$ is
opposite vertex $A$ in $G$, $B^*$ is opposite $B$, and $C^*$ is
opposite $C$, and each edge of $G^*$ crosses the appropriate edge of
$G$. (See Figure 2.) This results in a simultaneous embedding of
$L_{G^*, {\cal H}^*}$ and $L_{G, {\cal H}}$.  An example of the
construction for a particular generator is illustrated in Figure 3.

\begin{figure}
\includegraphics[scale=.33]{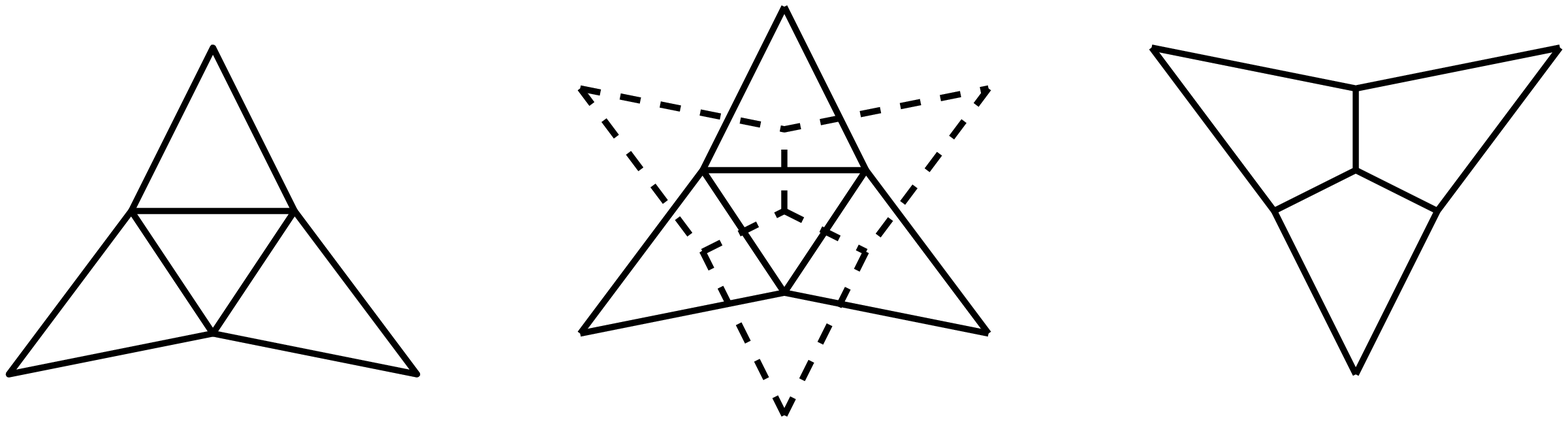}
\vspace{2mm}

\includegraphics[scale=.15]{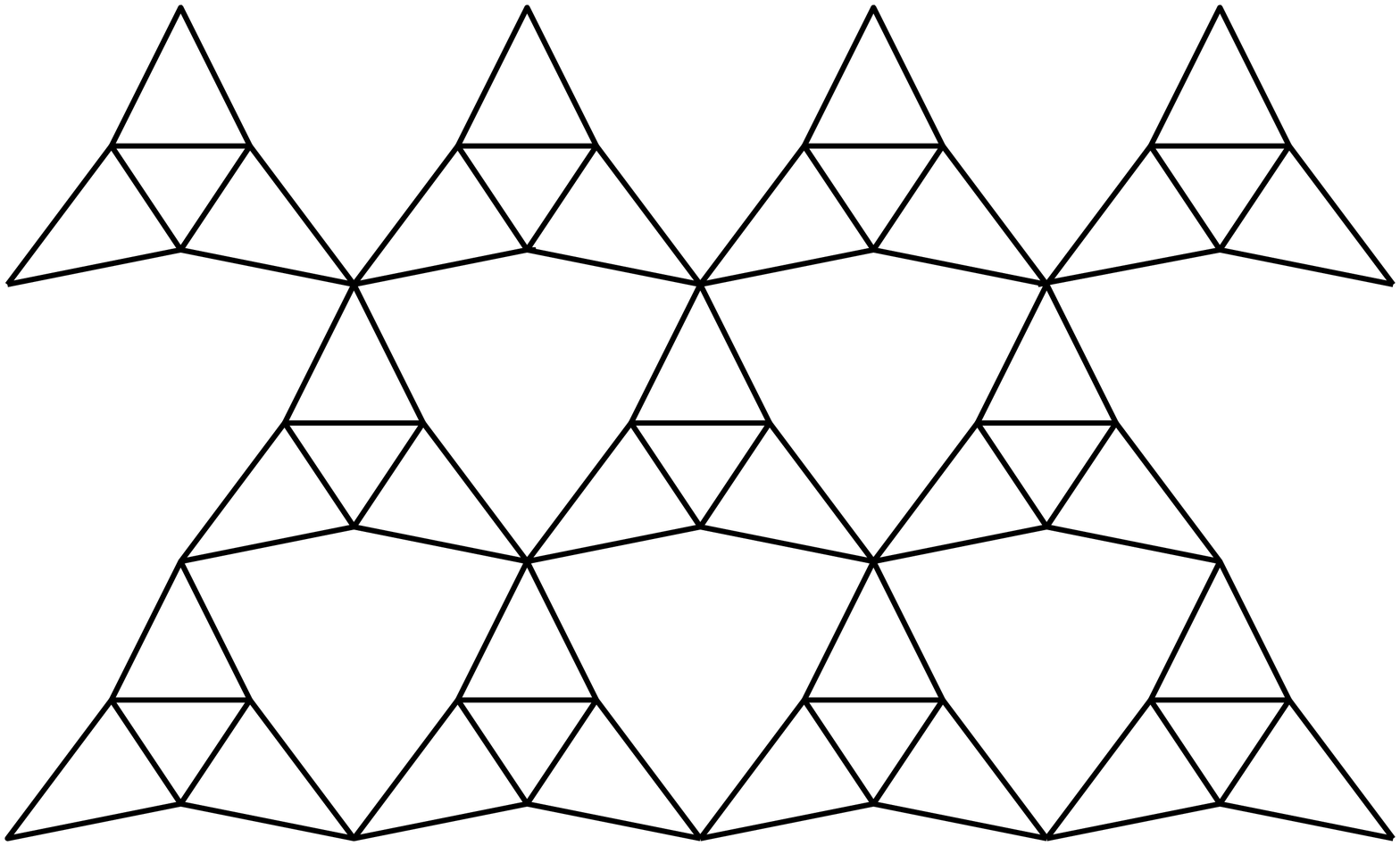}
\hspace{5mm}
\includegraphics[scale=.15]{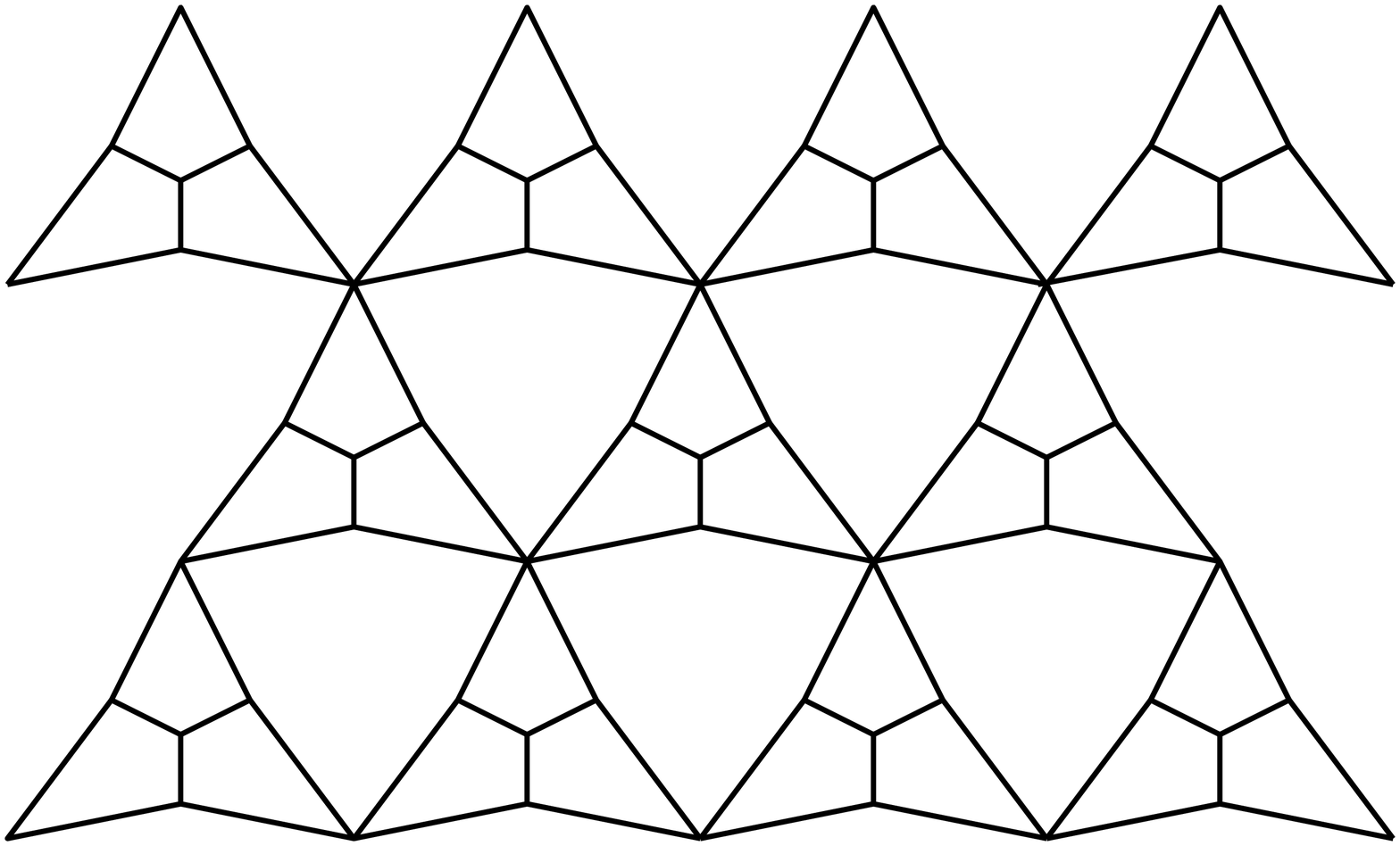}
\vspace{2mm}

\includegraphics[scale=.15]{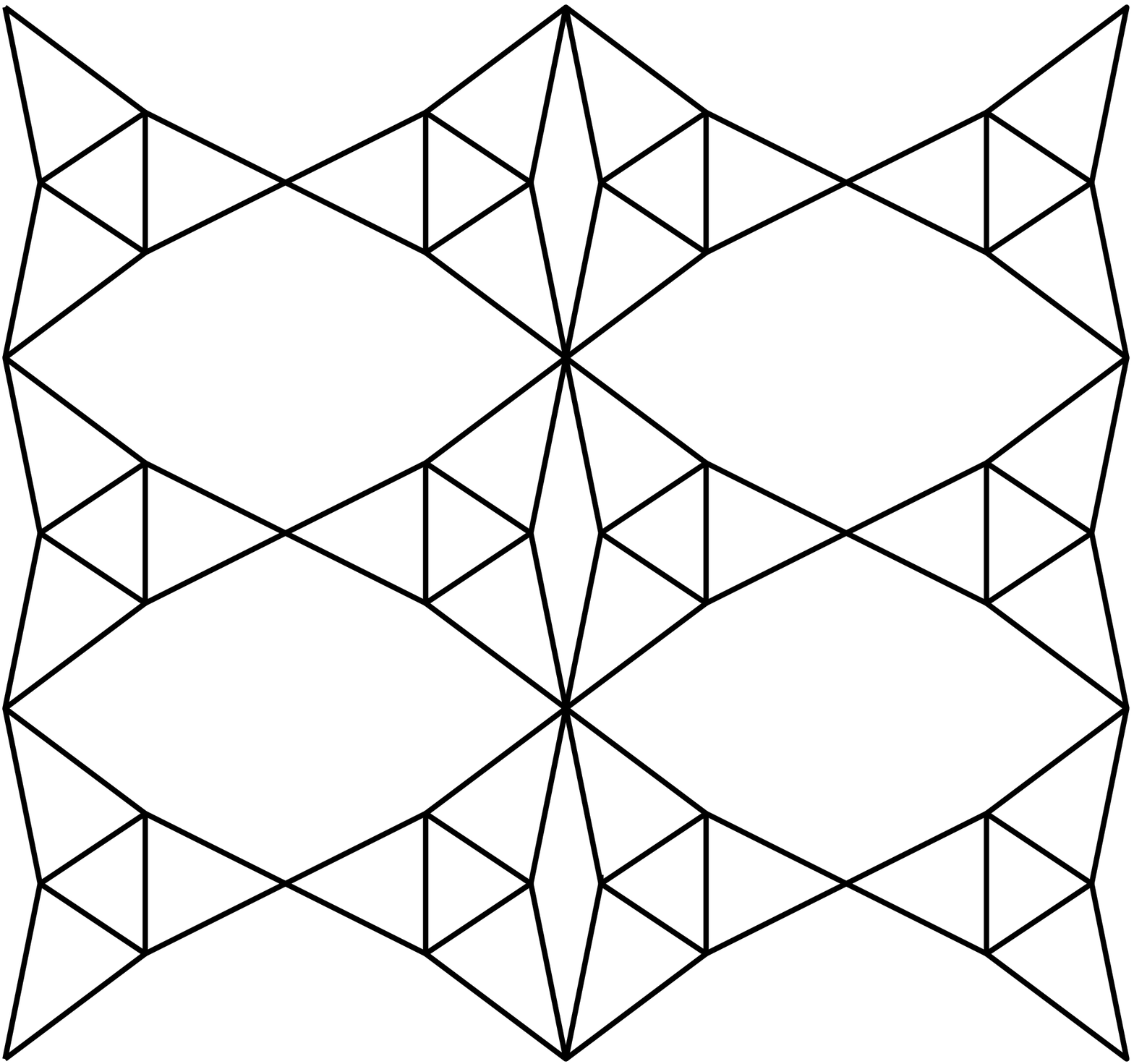}
\hspace{12mm}
\includegraphics[scale=.15]{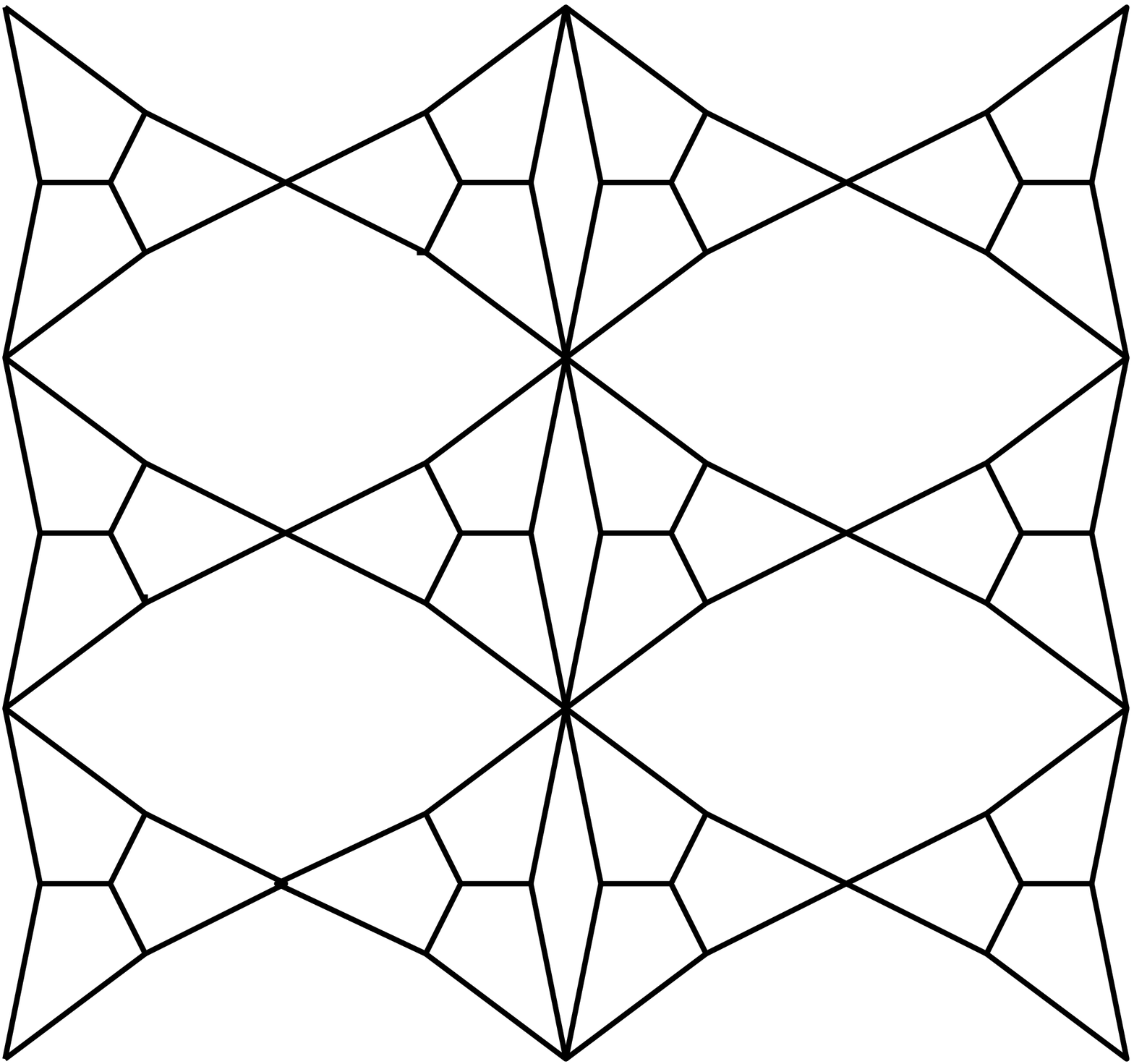}

\caption{The construction of lattices based on a specific generator.
Top: The generator, the duality relationship, and the dual
generator. Middle: The lattices based on the generator and the
triangular hypergraph arrangement.  Bottom: The lattices based on
the generator and the bow-tie hypergraph arrangement.}
\end{figure}

The constructions of the two lattices both produce a planar
representation of the resulting lattice. From the simultaneous
embeddings of the two lattices, it is seen that $L_{G^*, {\cal
H}^*}$ is the dual lattice of $L_{G, {\cal H}}$, since there is a
one-to-one correspondence between vertices of one and faces of the
other, and a one-to-one correspondence between edges, which are
paired by crossing.

\subsection{\label{sec:level2}Substitution Method}

Wierman \cite{Wie90} introduced the substitution method to find
bounds on bond percolation critical probabilities for certain
lattices.  Its application to the martini class of lattices is
described here. Let $\mathcal{G}$ be a lattice constructed by
placing copies of a generator in a self-dual 3-uniform hypergraph
arrangement. A boundary vertex is a vertex that is in more than one
copy of the generator. Let $G$ be the generator and suppose that its
$3$ boundary vertices are $A$, $B$, and $C$. Any configuration of
open and closed bonds on $G$ gives a partition of the boundary
vertices into connected open clusters of vertices. Vertical bars
between the boundary vertices are used to denote such a partition.
For example, $AB|C$ denotes the partition where $A$ and $B$ are in
the same open cluster and $C$ is in a separate open cluster.

\subsubsection{Partially Ordered Sets}

A {\it partially ordered set}, or briefly, a {\it poset}, consists
of a pair $(S, \leq)$, where $S$ is a set and $\leq$ is a binary
relation with the following properties: (1) $\leq$ is {\it
reflexive}, i.e., $s \leq s$ for all $s \in S.$ (2) $\leq$ is {\it
antisymmetric}, i.e., for all $s$ and $t$ in $S$, if $s \leq t$ and
$t \leq s$, then $s=t$. (3) $\leq$ is transitive, i.e., for all
$s$,$t$, and $u$ in $S$, if $s \leq t$ and $t \leq u$, then $s \leq
u$.

In a partially ordered set $(S, \leq)$, two elements of $S$, $s$ and
$t$, are {\it comparable} if $s \leq t$ or $t \leq s$, and are {\it
incomparable} otherwise. A partially ordered set is a {\it total
order} if every pair of its elements are comparable.  For example,
the standard relation $\leq$ on a set of real numbers is a total
order.  Partially ordered sets generalize the concept of total order
by allowing incomparability.  A common example of a partially
ordered set that is not totally ordered is the set of all subsets,
or {\it power set}, of a set $A$ of two or more elements with the
relation of set inclusion, since some pairs of subsets are
incomparable.

It is useful to have a visual representation of a partially ordered
set.  To efficiently represent a poset, it is useful to define the
{\it cover} relationship. For $s,t \in S$, $s$ is {\it covered by}
$t$ if $s \leq t$ and there is no element $u \in S$, unequal to both
$s$ and $t$, such that $s \leq u \leq t$.  The {\it Hasse diagram}
of the partially ordered set $(S, \leq)$ is obtained by placing a
point in the plane for each element of $S$, taking care to put the
point for $s$ below the point for $t$ whenever $s \leq t$, and
connecting the points for $s$ and $t$ by a line segment if and only
if $s$ is covered by $t$.  Note that all other comparability
relationships can be obtained from the covering relationships by
following monotone paths in the Hasse diagram, so it is not
necessary to complicate the diagram by including line segments for
all pairs of comparable elements.

The set of boundary partitions of a generator is a partially ordered
set, with, for boundary partitions $\pi$ and $\sigma$, $\pi \leq
\sigma$ whenever every cluster of $\sigma$ can be decomposed into
clusters of $\pi$. In the previous example, $A|B|C \leq AB|C$, since
$A|B$ is a decomposition of $AB$ and $C$ is a trivial decomposition
of $C$. Equivalently, every cluster of $\pi$ is entirely contained
in a single cluster of $\sigma$.  Another equivalent condition is
that every cluster of $\sigma$ is a disjoint union of clusters of
$\pi$.

\begin{figure}
\includegraphics[scale=.5]{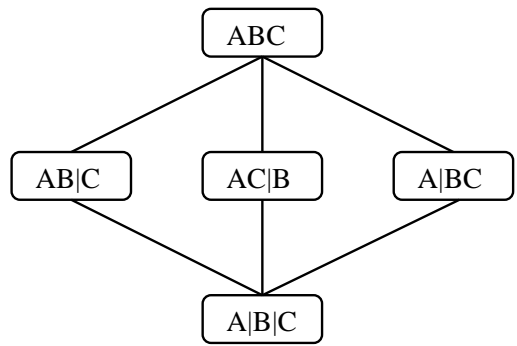}
\caption{A Hasse diagram of the partially ordered set of partitions
of three boundary vertices, ordered by refinement.}
\end{figure}

Figure 4 illustrates the Hasse diagram of a partially ordered set
with 5 elements which is used throughout this article. The three
middle partitions are pairwise incomparable, while all other pairs
of partitions are comparable.

\subsubsection{Stochastic Ordering}

Stochastic ordering is used to compare two probability measures
defined on the same partially ordered set.  We will consider
probability measures which are defined on the set of boundary
partitions of a generator in terms of a bond percolation model on
the generator.  Consider a generator $G$ and a bond percolation
model on $G$ with parameter $p$. A configuration is a designation of
each edge of $G$ as open or closed.  In the percolation model, a
configuration with $k$ open edges and $l$ closed edges has
probability $p^k(1-p)^l$.  A boundary partition is a union of
configurations, with its probability being the sum of the
probabilities of its configurations.  Let $P_p^G$ denote the
probability measure on boundary partitions of $G$ generated by the
bond percolation model with parameter $p$.

Let $S$ be a partially ordered set and let $U \subset S$. $U$ is
called an {\it upset} if for all $f$ and $g$ in $S$, $f \leq g$ and
$f \in U$ imply that $g \in U$.  If $P$ and $Q$ are two probability
measures on $S$, we say $P$ is stochastically smaller than $Q$,
denoted $P \leq_S Q$, if $P(U) \leq Q(U)$ for all upsets $U$ of $S$.
This concept of {\it stochastic ordering} is the appropriate
comparison of two probability measures on a partially ordered set
when applying the substitution method \cite{Wie92}.

In addition to a lattice $\mathcal{G}$, suppose another lattice
$\mathcal{H}$ is constructed by placing copies of a generator $H$ in
the same self-dual 3-uniform hypergraph structure as $\mathcal{G}$
The substitution method studies the effect on connection
probabilities of replacing the generator $H$ of $\mathcal{H}$ by the
generator $G$ in $\mathcal{G}$.

Preston \cite{Pre74} established the equivalence of stochastic
ordering and coupling for probability measures on finite partially
ordered sets. For simplicity here, we will specialize to our
percolation setting.  If $P_p^G \leq_S P_q^H$ for two generators $G$
and $H$, then there exist percolation models on $G$ and $H$ with
parameters $p$ and $q$ respectively which are dependent on each
other in such a manner that whenever boundary vertices are connected
by open edges in $G$ they are also connected by open edges in $H$.
(Note that each percolation model corresponds to independently open
edges in its generator, so is determined by its lattice
connectivity,  but that the realizations of the two percolation
models are stochastically dependent upon each other.) In particular,
the stochastic ordering $P_p^G \leq_S P_q^H$ implies that
$\theta^\mathcal{G}(p) \leq \theta^\mathcal{H}(q)$ and
$\chi^\mathcal{G}(p) \leq \chi^\mathcal{H}(q)$. We use these
consequences in section III, where solving for $\delta(\epsilon)$ in
the equations $P_{p_c+\epsilon}^L(U)=P_{q_c+\delta(\epsilon)}^D(U)$
for all upsets $U$ allows us to obtain a stochastic ordering on the
probability measures for a lattice and its dual near their
percolation thresholds.



\section{\label{sec:level1}Example: The Martini Lattice and Its Dual}

In this section, we show that the critical exponents of the bond
percolation models on the martini lattice and its dual, referred to
as the $K_4$ lattice, are equal. The generators for the martini and
$K_4$ lattices are shown in Figure 5. The method used is that
introduced by Wierman \cite{Wie92} to show the equality of the bond
percolation critical exponents for the triangular and hexagonal
lattices and also for the bowtie lattice and its dual.  There are no
other cases where Wierman's method has been used to show the
equality of bond percolation critical exponents for dual lattices.
This example is used as a reference in section IV, where the main
result is shown. Precisely, we show that $\beta^+(M) = \beta^+(K)$,
$\beta^-(M) = \beta^-(K)$, $\gamma^+(M) = \gamma^+(K)$, and
$\gamma^-(M) = \gamma^-(K)$, where $M$ and $K$ refer to the martini
and $K_4$ lattices respectively.   So, if the limit defining the
critical exponent $\beta$ exists for either lattice, it exists for
both lattices and $\beta(M)=\beta(K)$. Similarly, if the limit
defining the critical exponent $\gamma$ exists for either lattice,
it exists for both lattices and $\gamma(M)=\gamma(K)$. As in
Wierman's paper, the substitution method is used in this proof.

\begin{figure}[h]
        \includegraphics[scale=.40]{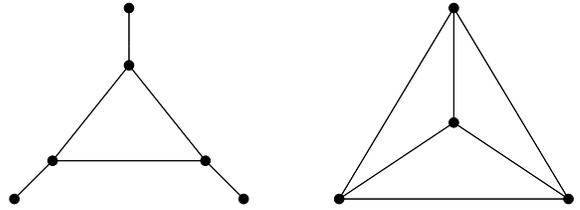}
        \caption{\label{1} Generators for the Martini and $K_4$ Lattices, on the left and right, respectively.}
\end{figure}

Ziff \cite{Zif06} was able to find the exact percolation threshold
for the martini and $K_4$ lattices.  A derivation of this threshold
is summarized here.  Let each bond in the $K_4$ lattice be open with
probability $p$.  Then, the partition probabilites for the $K_4$
lattice can be calculated by conditioning on the bonds in the
interior star:

\begin{center}
$P_p^K[ABC] = 3p^2 + 5p^3 -18p^4 + 15p^5 - 4p^6$
\begin{eqnarray*}
P_p^K[AB\vert C] &=& P_p^K[AC\vert B] = P_p^K[A\vert BC] \\
&=& p - p^2 - 5p^3 + 11p^4 - 8p^5 + 2p^6
\end{eqnarray*}
$P_p^K[A\vert B \vert C] = 1 - 3p + 10p^3 - 15p^4 + 9p^5 - 2p^6$ \\
\end{center}

\begin{figure}[h]
        \includegraphics[scale=.40]{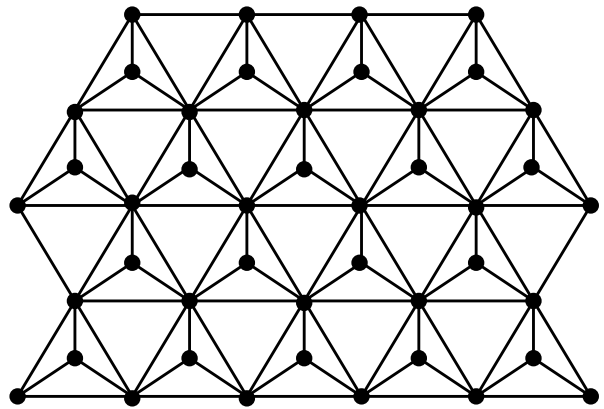}
    \caption{\label{2} The $K_4$ Lattice}
\end{figure}

Similarly, letting each bond be open in the martini lattice with
probability $q$, the partition probabilities for the martini lattice
can be calculated by conditioning on the bonds in the triangle:

\begin{center}
$P_q^M[ABC] = 3q^5 - 2q^6$
\begin{eqnarray*}
P_q^M[AB\vert C] &=& P_q^M[AC\vert B] = P_q^M[A\vert BC] \\
&=& q^3 + q^4 -4q^5 + 2q^6
\end{eqnarray*}
$P_q^M[A\vert B\vert C] = 1 - 3q^3 -3q^4 + 9q^5 - 4q^6$ \\
\end{center}

\begin{figure}[h]
        \includegraphics[scale=.35]{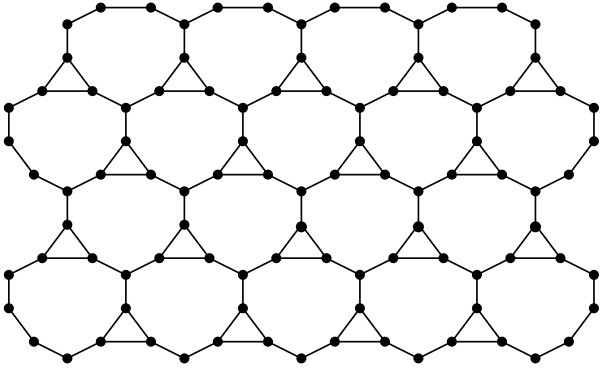}
    \caption{\label{2} The Martini Lattice}
\end{figure}

Using the substitution method, we replace each martini generator by
the $K_4$ generator and set $P_p^K[ABC]=P_q^M[ABC]$, $P_p^K[AB\vert
C]=P_q^M[AB\vert C]$, and $P_p^K[A\vert B \vert C]=P_q^M[A\vert
B\vert C]$.  The second equation is satisfied when $p=1-q$, which is
a consequence of duality of the two lattices.  The first and last
equations are redundant, and give the following:
$$(2q^2-1)(q^4-3q^3+2q^2+1)=0$$
which has roots at $q=\pm \frac{1}{\sqrt{2}}$.  We thus have
that the critical probabilities are $p_c(M)=\frac{1}{\sqrt{2}}$ and $p_c(K)=1-\frac{1}{\sqrt{2}}$
as shown before by Ziff \cite{Zif06}.\\

To show equality of the bond percolation critical exponents, we will
establish a stochastic ordering between the probability measures
$P_p^K$ near $p=p_c(K)$ and $P_q^M$ near $q=q_c(M)$. Specifically,
we find that
$$
P_{q_c+.93 \epsilon}^M \leq_S P_{p_c+\epsilon}^K \leq_S
P_{q_c+1.07\epsilon}^M
$$
by showing that
$$
P_{q_c+.93 \epsilon}^M (U) \leq P_{p_c+\epsilon}^K (U) \leq
P_{q_c+1.07\epsilon}^M (U)
$$
for all upsets $U$ of the poset of boundary partitions. This allows
us to make conclusions about the behavior of the percolation
probability near criticality.  Proceeding, we equate the upset
probabilities for the martini and $K_4$ lattices, keeping in mind
that they are equal at criticality as a result of duality. The upset
probability equations are given by:
\begin{center}
$3p^2 + 5p^3 - 18p^4 + 15p^5 - 4p^6 +$ \\
$k[p - p^2 - 5p^3 + 11p^4 - 8p^5 + 2p^6]$ \\
$= 3q^5 - 2q^6 + k[q^3 + q^4 - 4q^5 + 2q^6]$
\end{center}
for $k=0,1,2,3$, where $k$ denotes the number of middle partitions
in the upset. In studying critical exponents, we consider the
behavior of the system near the percolation threshold, so we
consider perturbations near $p_c$ and $q_c$.  Thus, we want to solve
for $\delta=\delta(\epsilon)$ in the following:
\begin{center}
$3(p_c+\epsilon)^2 + 5(p_c+\epsilon)^3 - 18(p_c+\epsilon)^4 +
15(p_c+\epsilon)^5 - 4(p_c+\epsilon)^6 $ $+k[(p_c+\epsilon) -
(p_c+\epsilon)^2 - 5(p_c+\epsilon)^3$ $ + 11(p_c+\epsilon)^4 -
8(p_c+\epsilon)^5 + 2(p_c+\epsilon)^6]$
\end{center}
\begin{center}
$= 3(q_c+\delta)^5 - 2(q_c+\delta)^6$ $+ k[(q_c+\delta)^3
+(q_c+\delta)^4 - 4(q_c+\delta)^5 + 2(q_c+\delta)^6],$
\end{center}
for $k=0,1,2,3$.

Noting that the constant terms cancel out by the upset probability equations and letting
\begin{center}
$g(p)=(6p + 15p^2 - 72p^3 + 75p^4 - 24p^5)$ \hspace{10mm}\\
\hspace{20mm} $+ k(1 - 2p - 15p^2 + 44p^3 - 40p^4 + 12p^5)$
\end{center}
and
$$h(q)=15q^4 + 12q^5 + k(3q^2 + 4q^3 - 20q^4 + 12q^5),$$
we see that
$$\delta(\epsilon) = \frac{g(p_c)}{h(q_c)}\epsilon + o(\epsilon).$$
Using the approximation $p_c = 1-.7071 = .2929$, the coefficients of $\epsilon$ are approximately:
\begin{center}
$1.0655$ for $k=0$\\
$1.0214$ for $k=1$\\
$.9791$ for $k=2$\\
$.9385$ for $k=3$\\
\end{center}

Thus, for sufficiently small $\epsilon$, $.93 \epsilon$ is smaller
than the coefficient in the solution $\delta(\epsilon)$ for all four
equations, and $1.07 \epsilon$ is larger than the coefficient in the
solution $\delta(\epsilon)$ for all four equations. Since the upset
probability functions are increasing functions of the parameters,
for sufficiently small positive $\epsilon$, we have
\begin{center}
$ 3(p_c+\epsilon)^2 + 5(p_c+\epsilon)^3 - 18(p_c+\epsilon)^4 +
15(p_c+\epsilon)^5 - 4(p_c+\epsilon)^6$ $+ k[(p_c+\epsilon) -
(p_c+\epsilon)^2 - 5(p_c+\epsilon)^3$ $+ 11(p_c+\epsilon)^4 -
8(p_c+\epsilon)^5 + 2(p_c+\epsilon)^6]$
\end{center}
\begin{center}
$\geq 3(q_c+.93 \epsilon)^5 - 2(q_c+.93 \epsilon)^6$ $ +k[(q_c+.93
\epsilon)^3 + (q_c+.93\epsilon)^4$ $ - 4(q_c+.93 \epsilon)^5 +
2(q_c+.93 \epsilon)^6],$
\end{center}
for $k=0,1,2,3$, and

\begin{center} $3(p_c+\epsilon)^2 +
5(p_c+\epsilon)^3 - 18(p_c+\epsilon)^4 + 15(p_c+\epsilon)^5 -
4(p_c+\epsilon)^6 $ $+k[(p_c+\epsilon) - (p_c+\epsilon)^2 -
5(p_c+\epsilon)^3$ $ + 11(p_c+\epsilon)^4 - 8(p_c+\epsilon)^5 +
2(p_c+\epsilon)^6]$
\end{center}
\begin{center}
 $\leq 3(q_c+1.07\epsilon)^5 -
2(q_c+1.07\epsilon)^6$ $+ k[(q_c+1.07\epsilon)^3 +
(q_c+1.07\epsilon)^4$ $ - 4(q_c+1.07\epsilon)^5 +
2(q_c+1.07\epsilon)^6],$
\end{center}
for $k=0,1,2,3$.

Each of these sets of inequalities establishes a stochastic ordering
result. Thus, for sufficiently small $\epsilon
> 0$,
$$P^M_{q_c+.93\epsilon} \leq_S P^K_{p_c+\epsilon} \leq_S P^M_{q_c+1.07\epsilon}$$

Using the result from Preston \cite{Pre74}, we have that, for $p>p_c(K)$ and sufficiently close to $p_c(M)$,
$$\theta^M(q_c+.93(p-p_c)) \leq \theta^K(p) \leq \theta^M(q_c + 1.07(p-p_c))$$
Taking logarithms and dividing by $\log|p-p_c|$ throughout, we see that 
\begin{eqnarray*}
\mathop {\limsup }\limits_{q \downarrow q_c} \frac{\log \theta^M(q)}{\log|q-q_c|} &=&
\mathop {\limsup }\limits_{p \downarrow p_c} \frac{\log \theta^M(q_c+.93(p-p_c))}{\log|.93(p-p_c)|}\\
&=& \mathop {\limsup }\limits_{p \downarrow p_c} \frac{\log \theta^M(q_c+.93(p-p_c))}{\log(.93) + \log|p-p_c|} \\
&=& \mathop {\limsup }\limits_{p \downarrow p_c} \frac{\log \theta^M(q_c+.93(p-p_c))}{\log|p-p_c|} \\
&\geq& \mathop {\limsup }\limits_{p \downarrow p_c} \frac{\log \theta^K(p)}{\log|p-p_c|}
\end{eqnarray*}

Reversing this argument using $q=q_c+1.07(p-p_c)$ gives:
$$ \mathop {\limsup }\limits_{p \downarrow p_c} \frac{\log \theta^K(p)}{\log|p-p_c|}
\geq \mathop {\limsup }\limits_{q \downarrow q_c} \frac{\log \theta^M(q)}{\log|q-q_c|}$$

Consequently, $\beta^+(K)=\beta^+(M)$ and, by changing limsup to liminf where it appears above,
we have that $\beta^-(K)=\beta^-(M)$. \\

By the same reasoning, we also have that for sufficiently small $\epsilon>0$,
$$P^M_{q_c-1.07\epsilon}\leq_S P^K_{p_c-\epsilon} \leq P^M_{q_c-.93\epsilon}$$
Thus, for $p<p_c(K)$ with $p$ sufficiently close to $p_c(M)$,
$$\chi^M(q_c-1.07\epsilon) \leq \chi^K(p_c-\epsilon) \leq \chi^M(q_c-.93\epsilon).$$
Letting $\epsilon \downarrow 0$ gives that $\gamma^+(K)=\gamma^+(M)$ and $\gamma^-(K)=\gamma^-(M)$.

\section{\label{sec:level1}Positive Coefficients for
$\epsilon$ in the $\delta(\epsilon)$ functions}

In this section we generalize the result to lattices constructed
from an infinite connected planar periodic 3-uniform hypergraph with
one axis of symmetry, using a generator which is a finite connected
planar graph with three boundary vertices.   See Wierman and Ziff
\cite{Wie08} for the construction of such lattices.  As in section
III, the method used to show equality of the critical exponents is
valid if the coefficients of $\epsilon$ in the $\delta(\epsilon)$
functions are positive and finite.  In this section we show that the
coefficients for lattices in the  class are positive and finite. In
what follows, $G$ will denote the generator of a lattice in the
class of interest
and $D$ will denote its dual generator.\\

Calculate the coefficient of $\epsilon$ in the $\delta(\epsilon)$
function as follows.  First, equate the upset probabilites for the
lattice and its dual, setting
\begin{center}
$P_p^G[ABC] + jP_p^G[AB|C] + kP_p^G[AC|B] + lP_p^G[BC|A] =
P_q^D[ABC] + jP_q^D[AB|C] + kP_q^D[AC|B] + lP_q^D[BC|A]$,
\end{center}
where $j$, $k$, and $l$ take the values $0$ or $1$.
Note that this is more general than the martini and $K_4$
lattices example in that symmetry of the generator is not assumed.
The critical probabilities for the lattice and its dual are
then calculated using these equations.  Adding $\epsilon$ to
$p$ and $\delta$ to $q$ in these equations shows how the
two functions behave around the critical probabilities. \\

Notice that the numerator of the coefficent of $\epsilon$ in the
martini and $K_4$ lattices example is the derivative of
$P_p^K[ABC]+kP_p^K[A|BC]$ and the denominator is the derivative of
$P_q^M[ABC]+kP_q^M[A|BC]$.  This holds for all generators in the
class. That is, the numerator of the coefficent of $\epsilon$ is the
derivative of $P_p^G[ABC]+jP_p^G[A|BC]+kP_p^G[AB|C]+lP_p^G[AC|B]$
and the denominator is the derivative of
$P_q^D[ABC]+jP_q^D[A|BC]+kP_q^D[AB|C]+lP_q^D[AC|B]$ for any
generator in the class.  This can be seen by the following
reasoning: Suppose the generator $G$ has $n$ bonds. Then, since the
lattice's generator and its dual have the same number of bonds, the
expressions $P_p^G[ABC]+jP_p^G[A|BC]+kP_p^G[AB|C]+lP_p^G[AC|B]$ and
$P_q^D[ABC]+jP_q^D[A|BC]+kP_q^D[AB|C]+lP_q^D[AC|B]$ are polynomials
in $p$ and $q$ respectively with degree
no larger than $n$.  So, we can write \\

$P_p^G[ABC]+jP_p^G[A|BC]+kP_p^G[AB|C]+lP_p^G[AC|B]$
$$= \sum_{k=0}^n a_kp^k = g(p)$$
and
$P_q^D[ABC]+jP_q^D[A|BC]+kP_q^D[AB|C]+lP_q^D[AC|B]$
$$= \sum_{k=0}^n b_kq^k = h(q)$$

Equating these, and adding $\epsilon$ and $\delta$ to $p$ and $q$ respectively,
yield the following:
$$\sum_{k=0}^n b_k(q+\delta)^k = \sum_{k=0}^n a_k(p+\epsilon)^k.$$
Using binomial expansions, we can write this as:
$$\sum_{k=0}^n b_kq^k + \delta \sum_{k=1}^n kb_kq^{k-1} +
o(\delta)$$ \vspace{-5mm}
$$= \sum_{k=0}^n a_kp^k + \epsilon \sum_{k=1}^n
ka_kp^{k-1} + o(\epsilon)$$

Evaluating this at $p=p_c$ and $q=q_c$, the constant terms on each
side of this equation cancel as a result of the equality of the
probability measures for dual lattices at criticality.  Moreover,
the coefficients of $\delta$ and $\epsilon$ are seen to be the
derivatives of their respective upset probability functions.  We
thus have that
$$\delta(\epsilon) = \frac{\frac{d}{dp}g(p)|_{p_c}}{\frac{d}{dq}h(q)|_{q_c}}\epsilon + o(\epsilon)$$

To show that the coefficient of epsilon is positive and finite,
it suffices to show that the derivatives of the upset probability
functions in both the lattice and its dual are positive on the
interval (0,1).  Since the upset probability functions are polynomials,
their derivatives exist and are finite.  If the derivatives are shown
to be positive, then they will be positive at $p_c$ and the coefficient
of $\epsilon$ will be positive, completing the argument. \\

We now prove that the derivatives of the partition probability
functions for a lattice in the martini class and its dual are
positive. Let $T$ be a minimal connected subgraph of the generator
$G$ that contains $A$, $B$, and $C$.  Let $E(T)$ denote the edge set
of $T$ and let $m=|E(T)|$.  $T$ is clearly a tree, so there is an
unique path from $A$ to $B$, from $A$ to $C$, and from $B$ to $C$.
More importantly, removing any edge from $T$ makes connectivity
between $A$, $B$, and $C$ impossible in $T$.

We shall use the following notation for the proof.  The superscript
on the probability measure will indicate the structure for which the
boundary vertices can be connected through; $G$ will indicate the
generator of the lattice, $T$ will indicate the tree defined above,
and $N$ will indicate through $G$ but not through $T$. (The meaning
of this will become clearer in the body of the proof.) Where no
superscript appears, the statement is true for all three structures.
The subscript will denote the probability of an edge being open in
the graph.  Where two subscripts appear, separated by a comma, the
first subscript gives the probability that $e_1$ is open, and the
second subscript gives the probability that each edge other than
$e_1$ is open. $e_1$ may be defined differently in different cases,
but $e_1$ will always be an edge in $T$.  Where no subscript
appears, the statement is true for any subscript.  It will be
helpful to condition on $e_1$ being open or closed.  A semicolon
will separate the upset from the conditioning event.  $P(U;e_1)$
denotes the probability of the upset $U$ given that the edge $e_1$
is open, while $P(U;\overline{e_1})$ denotes the probability of the
upset $U$ given that the edge $e_1$ is closed.

Using this notation and continuing the discussion preceding the
notational description, $P_p^T[ABC;\overline{e_1}]=0$.

Let each bond in a lattice be open with probability $p$, $0<p<1$.
Let $e_1$ be an edge in $E(T)$, and therefore in $E(G)$, and let
$e_2$, $e_3$, \ldots, $e_n$ be the other bonds in $G$.  Let
$\{X_j\}_{j=1}^n$ be independent uniform random variables on the
interval $(0,1)$. For each realization $\overrightarrow{x}$ of the
$\{X_j\}_{j=1}^n$, if $x_j < p$, call $e_j$ open.  Otherwise, $e_j$
is closed. Call this scenerio Model 1.  Note that $P_p^G[ABC]$
denotes the probability that $A$, $B$, and $C$ are all connected in
$G$ in this case.\\

Consider the same lattice with the only difference being
that $e_1$ is open with probability $p+\epsilon$.
Call this scenerio Model 2, and note that $P_{p+\epsilon,p}[ABC]$
denotes the probability that A, B, and C are all connected in this case.
The other partition probability functions are defined in this
case using the same subscript. \\

Call Model 3 the same as Model 1 with $p$ replaced by $p+\epsilon$
in both the description and the notation.  We use the same
realization $\overrightarrow{x}$ in all three cases.\\

Notice that if an edge is open in Model 1, it is also open in Model 2.
So, if $A$, $B$, and $C$ are connected through open bonds in Model 1,
they are necessarily connected through open bonds in Model 2.
Also, if an edge is open in Model 2, it is open in Model 3 and
similar conclusions can be made. \\

In what follows, $P^T[ABC]$ is the probability that $A$, $B$, and
$C$ are connected in $G$ only through $T$, that is, they are not
connected if any edge of $T$ is closed. $P^N[ABC]$ is the
probability that $A$, $B$, and $C$ are connected in $G - e$, the
graph $G$ with edge $e$ deleted, for some edge $e \in E(T).$ We then
have that $P^G[ABC] = P^T[ABC] + P^N[ABC]$. We obtain a lower bound
on the derivative by considering the difference quotients: \\

$\mathop {\lim }\limits_{\epsilon \to 0} \frac{P_{p+\epsilon}^G[ABC] - P_p^G[ABC]}{\epsilon}$ \\
$\geq \mathop {\lim }\limits_{\epsilon \to 0} \frac{P_{p+\epsilon,p}^G[ABC] - P_p^G[ABC]}{\epsilon}\\
= \mathop {\lim }\limits_{\epsilon \to 0} \frac{P_{p+\epsilon,p}^T[ABC]
+ P_{p+\epsilon,p}^N[ABC] - (P_p^T[ABC] + P_p^N[ABC])}{\epsilon}$\\

Substituting
$$P_{p+\epsilon,p}[ABC]=(p+\epsilon)P_p[ABC;e_1] + (1-p-\epsilon)P_p[ABC;\overline{e_1}]$$
$$ \text{and } P_p[ABC]=pP_p[ABC;e_1] + (1-p)P_p[ABC;\overline{e_1}],$$
we see that the last expression above equals:
$$P_p^T[ABC;e_1] - P_p^T[ABC;\overline{e_1}]+ P_p^N[ABC;e_1] - P_p^N[ABC;\overline{e_1}]$$

Since $P_p^T[ABC;e_1]=p^{m-1}$, $P_p^T[ABC;\overline{e_1}]=0$, and
the event that $A$, $B$, and $C$ are connected in $N$ with $e_1$
closed is contained in the event that the three are connected,
$$\mathop {\lim }\limits_{\epsilon \to 0} \frac{P_{p+\epsilon}^G[ABC]
- P_p^G[ABC]}{\epsilon} \geq p^{m-1} + 0 > 0$$ Similar reasoning is
valid for the other partition probability functions, one of which
will be given in detail here. Let $P[AB] = P[ABC] + P[AB|C]$, the
probability that $A$ and $B$ are connected in the lattice.  Also,
let $e_1$ be an edge in $T$ on the unique path from $A$ to $B$ and
let $a$ be the number of edges in the path.  Then, \\

$\mathop {\lim }\limits_{\epsilon \to 0} \frac{P_{p+\epsilon}^G[AB] - P_p^G[AB]}{\epsilon}$ \\
$\geq \mathop {\lim }\limits_{\epsilon \to 0} \frac{P_{p+\epsilon,p}^G[AB] - P_p^G[AB]}{\epsilon}\\
= \mathop {\lim }\limits_{\epsilon \to 0} \frac{P_{p+\epsilon,p}^T[AB]
+ P_{p+\epsilon,p}^N[AB] - (P_p^T[AB] + P_p^N[AB])}{\epsilon}\\
=\mathop {\lim }\limits_{\epsilon \to 0} \frac{\epsilon
(P_p^T[AB;e_1] + P_p^N[AB;e_1]
+ P_p^T[AB;\overline{e_1}] +  P_p^N[AB;\overline{e_1}])}{\epsilon} \\
= P_p^T[AB;e_1] + (P_p^N[AB;e_1] + P_p^N[AB;\overline{e_1}]) \\
\geq p^{a-1} + 0>0$, \\

In this way, it follows that all upset probability functions for a
generator of a lattice in the class have a positive derivative in
the interval (0,1).  The exact same argument holds for the duals of
these lattices, replacing $p$ by $q$.  Thus, the coefficient of
$\epsilon$ used in this method to determine equality of the bond
percolation critical exponents is well-defined and positive.  As a
consequence, any lattice and dual lattice in the class described
have equal values for the two critical exponents $\beta$ and
$\gamma$.

\section{\label{sec:level1}Extensions and Generalizations}

\subsection{\label{sec:level2}Other Critical Exponents}

Additional percolation functions and critical exponents are considered in the literature.
Define the two-point connectivity function by:\\
$\tau(p,x,y)=P_p[x \leftrightarrow y]$, where ``$x \leftrightarrow
y$" denotes the event that the sites $x$ and $y$ are connected by an
open path.

Define the correlation length by
$$\xi(p) = \left[ \frac{1}{\chi_f(p)} \sum_y |y-v|^2 P_p[v \to y,|C_v|<\infty]]\right]^\frac{1}{2},$$
where $|y|=\max \{|y(i)|:1 \leq i \leq d \}$,
$y=(y(1),\ldots,y(d)).$

The critical exponents $\nu$, $\delta$, $\eta$, and $\Delta_k$,
for $k \geq 2$, are given by the power laws: \\

$\xi(p) \approx |p-p_c|^{-\nu}$ for some $\nu > 0$, \\

$\frac{E_p[|C|^k;|C|<\infty]}{E_p[|C|^{k-1};|C|<\infty]} \approx |p-p_c|^{-\Delta_k}$, $k \geq 2$,
for some $\Delta_k>0$, \\

$P_{p_c}[n\leq |C|<\infty] \approx n^{1/\delta}$, $n \to \infty$, for some $\delta > 0$, \\

\noindent and, for a $d$-dimensional periodic graph, \\

$\tau(p_c,v,x) \approx |x|^{2-d-\eta}$, $|x| \to \infty$, for some $\eta>0$. \\

In two dimensions, the values for the critical exponents defined in
this paper are believed to be $\beta=5/36$, $\gamma=43/18$,
$\delta=91/5$, and $\nu=4/3$ (See Grimmett \cite{Gri99} and Hughes
\cite{Hug96}).  Kesten \cite{Kes87} proved that for a class of
two-dimensional periodic lattices (assuming the limits defining the
exponents $\delta$ and $\nu$ exist), that
$$\beta=\frac{2\nu}{\delta+1}, \gamma=2\nu \frac{\delta-1}{\delta+1}, \eta=\frac{4}{\delta+1}$$
and that for all $k \geq 2$,
$$\Delta = \Delta_k = 2\nu \frac{\delta}{\delta+1}.$$
The class of lattices considered by Kesten includes the lattices in
the class identified by Wierman and Ziff.

We have already shown that $\beta(L)=\beta(D)$ and
$\gamma(L)=\gamma(D)$ for any lattice $L$ in the given class.
Furthermore, since $P_{p_c}^L$ and $P_{q_c}^D$ are equal,
$\tau(q_c,x,y)=\tau(p_c,x,y)$ for all boundary vertices $x$ and $y$.
Thus, $\tau(q_c,x,y)$ and $\tau(p_c,x,y)$ decrease at the same
exponential rate, so $\eta(K)=\eta(M)$.  Using Kesten's formulas,
equality of the other critical exponents is evident.  That is,
$\delta(L)=\delta(D)$, $\nu(L)=\nu(D)$, and $\Delta(L)=\Delta(D)$.

Thus, the set of critical exponents are equal for a bond model and
its dual in the given class.  For example, the results apply to the
lattice pairs mentioned in Ziff and Scullard \cite{Zif062}. Since
each generator can appear in an infinite collection of self-dual
hypergraphs, we have infinitely many lattices in which the set of
critical exponents are equal.

Our results do not establish any numerical values for the critical
exponents.  However, remarkable progress in this direction was made
by Smirnov and Werner \cite{SmWe}, who combined Kesten's scaling
relations, knowledge of critical exponents associated with the
stochastic Loewner evolution process, and Smirnov's proof of
conformal invariance to determine the existence and values of
critical exponents for the site percolation model on the triangular
lattice.

\subsection{\label{sec:level2}The Bowtie Lattice and Its Dual}

By splitting each vertical bond in the bowtie lattice into two
bonds, each having probability $1-\sqrt{1-p}$ of being open, Wierman
\cite{Wie92} used the substitution method to determine the critical
probabilities of the bowtie lattice and its dual.  The method
described in this paper was then applied to this dual pair, showing
the equality of their bond percolation critical exponents. Using
this idea, the method described in this paper is applicable to many
other lattices.  In fact, as long as a lattice is self-dual under
the triangle-triangle transformation (see Ziff and Scullard
\cite{Zif062}, and Wierman and Ziff \cite{Wie08}), functions of $p$
can be assigned as edge probabilities of a given lattice.  That is,
we can assign to each edge of the generator the probability $f_e(p)$
of being open, where $f_e(p)$ is an increasing, right-continuous
function of $p$.  Under these conditions, repeating the main
argument of the paper gives that, for any edge $e$ in the minimal
tree connecting $A$, $B$, and $C$,
$$\mathop {\lim }\limits_{\epsilon \to 0} \frac{P_{p+\epsilon}^G[ABC] - P_p^G[ABC]}{\epsilon}$$
$$\geq f_e'(p)(P_p^T[ABC] + P_p^N[ABC|e]-P_p^N[ABC|\overline{e}])$$
and it is clear that the coefficient of $\epsilon$ in the expression
for $\delta(\epsilon)$ is well-defined and positive.  Equality of
bond percolation critical exponents for such lattices and their
duals follows from this fact.

\subsection{\label{sec:level2}Site Percolation}

Since the bond problem of a lattice is equivalent to the site
problem on its line lattice \cite{Kes82,Gri99,Fis61,Fis612}, the
result described in this paper applies to the collection of site
problems obtained by reformulating the bond problems as site
problems.  As a result, we have a collection of site problems in
which any lattice belonging to the collection shares the same value
of its critical exponents with its matching lattice.  So, for the
bond model on any given lattice in the class identified by Wierman
and Ziff, we can immediately identify three other lattices with the
same values for the set of critical exponents: the bond model on the
lattice's dual, the corresponding site problem on the line lattice,
and the site problem on the line lattice of the dual. Since each
generator can appear in an infinite collection of self-dual
hypergraphs, we have infinitely many site problems in which the set
of critical exponents are equal.

The martini, A, and B lattices discussed in Scullard \cite{Scu06}
are not line graphs of underlying lattices, so the results in the
previous paragraph do not apply to the site percolation models on
these lattices. It is plausible that the approach of this article
may be applicable to such site percolation models, but that has not
yet been shown to be valid, and is a subject of further study.

\section{\label{sec:level1}Concluding Remarks}

The method used by Wierman to show equality of the bond percolation
critical exponents for the triangle and hexagonal lattices was used
to show the equality of the bond percolation critical exponents for
the martini lattice and its dual.  This computational proof was then
extended to show that a lattice constructed from an infinite
connected planar periodic 3-uniform hypergraph with one axis of
symmetry, using a generator which is a finite connected planar graph
with three boundary vertices, has the same values for its bond
critical exponents as its dual, thereby generalizing the results of
Wierman \cite{Wie92}. Since the mentioned class of lattices is
infinite, there are infinitely many lattices that have the same bond
percolation critical exponents as their duals.  Moreover, since
using the same generator on different self-dual hypergraphs does not
affect the computations, there are an infinite number of bond and
site models for which the critical exponents are the same.  This
result gives mathematical evidence that the values of the critical
exponents may only depend on the dimension of the lattice,
supporting the universality hypothesis.  Note that the result does
not say that the bond percolation critical exponents have the same
value for all lattices in the mentioned class.  Using different
generators produces different upset probability functions, so no
relations for the bond percolation critical exponents between
generators has been determined.  If two non-isomorphic, non-dual
generators were discovered that had equal upset probability
functions at criticality, the result could likely be extended to
give equality of bond percolation critical exponents for these
lattices.  The authors of this paper have thus far been unable to
identify two such generators.

\end{document}